# Temporal characteristics and energy deposition of pulsating auroral patches

B. K. Humberset[1], J. W. Gjerloev[1,2], M. Samara[3], R. G. Michell[3,4], and I. R. Mann[5]

[1]Birkeland Centre for Space Science, Department of Physics and Technology, University of Bergen, Bergen, Norway, [2]The Johns Hopkins University Applied Physics Laboratory, Laurel, Maryland, USA, [3]NASA Goddard Space Flight Center, Greenbelt, Maryland, USA, [4]Department of Astronomy, University of Maryland, College Park, Maryland, USA, [5]Department of Physics, University of Alberta, Edmonton, Alberta, Canada

**Abstract** We present a careful statistical analysis of pulsating aurora (PA) using all-sky green line (557.7 nm) images obtained at 3.3 Hz. Six well-defined individual PA patches are identified and extracted using a contouring technique. Quantitative parameters such as the patch duration (on-time and off-time), peak intensity, and integrated intensity are determined for each patch and each pulsation. The resulting characteristics serve as strict observational constraints that any of the many competing theories attempting to explain PA must predict. The purpose of this paper is to determine the characteristics of PA patches in order to provide better observational constraints on the suggested mechanisms. All aspects of the temporal behavior of the individual patches appear to be erratic. Historically, PA has been defined very loosely and we argue that the use of the term "pulsating" is inappropriate since our findings and other published results are not regularly periodic and thus a more appropriate term may be fluctuating aurora. Further, we find that the observational constraints do not fit well with the flow cyclotron maser theory, which in particular is suggested to create PA patches. There is no clear candidate of the suggested mechanisms and drivers to explain the observational constraints set by the PA patches in a satisfactory manner.

## 1. Introduction

Pulsating aurora (PA) is a phenomenon of irregularly shaped patches and bands of low-intensity aurora that undergoes rapid alternating increases and decreases in luminosity. Patches of different sizes and shapes switch on and off or vary in intensity independently of each other. The broad definition of pulsating aurora covers auroral arcs, arc segments, and patches of fixed and variable area having horizontal sizes of a few to hundreds of kilometers with repetitive, quasiperiodic, or occasionally, periodic intensity variations on time scales ranging from less than 1 s to several tens of seconds [*Royrvik and Davis*, 1977].

The pulsating aurora is generally considered to be quasiperiodic with an average period of 8 ± 2 s [*Royrvik and Davis*, 1977]. It is called quasiperiodic because in a single train of pulsations the spacing between successive maxima is not even but varies noticeably from one pulsation to the next. The range 2–20 s has been chosen because observations falling within the chosen definition apparently form a uniform data set and are concerned with a single magnetospheric process, while there are temporal changes in the aurora with both shorter and longer periods which seem to have different characteristics [*Johnstone*, 1978]. The light intensity is often described having rapid rise and decay times compared to the duration of the pulse, as if it were switching on and off, in addition to the sometimes simultaneous ~3 Hz pulsations on top. In general, there is a characteristic variety in the shape of a train of pulsations. Both "on-time" and "off-time" (see Figure 6) can vary from pulse to pulse, with the larger variations probably found in the off-time [*Davidson and Chiu*, 1991]. Attempts have been made to relate pulsation periods to latitude. *Thomas and Rothwell* [1979] found a relation interpreted as being consistent with the latitude variation of the bounce period. However, in a later study, *Duncan et al.* [1981] found no consistent, simultaneous change in period with latitude, which have discouraged further investigations of a connection to bounce period or Alfvén communication. *Royrvik and Davis* [1977] were unable to find any significant trends that relate pulsating behavior to form type, level of activity, time within the auroral substorm, local time, or latitude. It seem like this conclusion still stands, although there are indications that PA with shorter periodicity can have a near-Earth origin [e.g., *Johnstone*, 1971], while longer periodicity PA is linked to a magnetospheric source [e.g., *Jaynes et al.*, 2013].

There is an agreement that the high-energy precipitating electrons result from pitch angle scattering. However, exactly how the modulation causing the pulsating aurora occurs is still up for discussion. Most studies point toward a time-varying pitch angle scattering close to the magnetic equator. In situ studies have observed lower band chorus [*Nishimura et al.*, 2010] and continuous measurements of electron flux modulations [*Jaynes et al.*, 2013] correlated with a single pulsating patch located through cross-correlation analysis of image pixels from an all-sky imager having the mapped footpoint within its field of view. The nonlinear relaxation oscillator [*Davidson*, 1979, 1986a, 1986b] and the flow cyclotron maser [*Demekhov and Trakhtengerts*, 1994] are theoretical candidates explaining the time-varying pitch angle scattering, the latter theory in a more quantitative manner which makes it easier to compare with observations. Since then, an increasing number of observations at the magnetic equator have made it possible to compare local plasma parameters (chorus waves, electron cyclotron harmonic waves, plasma densities, and ULF waves) in order to understand the time-varying pitch angle scattering. For example, depletions in the total electron density, probably attributed to changes in cold electron fluxes of unknown origin, are found to correlate with increases in chorus wave amplitude [*Li et al.*, 2012]. This kind of studies will probably make new advances in understanding the processes behind pulsating aurora, but there are also well-known characteristics such as nonconjugate pulsating aurora that are hard to explain with a single mechanism at the magnetic equator. *Sato et al.* [2002, 2004] showed observations of velocity-dispersed high-energy electron precipitation in correlation with nonconjugate pulsating aurora and anticorrelated with the proton flux, suggesting that it was due to a time-varying field-aligned electric field far from the magnetic equator and possibly a good fit to the mechanism of auroral acceleration region modulation by Alfvén waves [*Fedorov et al.*, 2004]. There seems to be an agreement that the ionosphere is not entirely passive, but exactly what role it has is largely unclear. There are not many suggestions of a source local to lower altitudes, such as the concept of how neutral atmosphere pressure waves could cause quasiperiodic fluctuations in auroral intensity described by *Luhmann* [1979]. It is more likely that the lower ionosphere and atmosphere have an alternating component, such as an ionospheric feedback mechanism based on the flow cyclotron maser theory suggested by *Tagirov et al.* [1999]. Pulsating aurora is more complicated than can probably be explained by time-varying pitch angle scattering alone. It remains to find how one or more ionospheric or near-Earth mechanisms contribute, if they act as a secondary control mechanism on the time-varying electron flux scattered near the magnetic equator or if they are entirely responsible for some aurora that falls within the broad category of pulsating aurora.

A satellite at low-Earth orbit altitudes crosses a pulsating patch in less than a pulsating period, while a slower-moving rocket would cross the patch within a few on-off cycles. Within that time, the observed variation is likely both temporal and spatial, and the sampling rate is limited. For high-altitude satellites (e.g., geostationary) the magnetic field line mapping is very challenging and thus we must question the relationship between the auroral form and the satellite observations.

The purpose of this paper is to determine the characteristics of PA patches in order to provide better observational constraints on the suggested mechanisms. We use ground-based all-sky imager observations which provide good spatial and temporal resolution of a PA patch. The two-dimensional images of the night sky allow us to objectively separate spatial and temporal variations, thereby avoiding the space-time ambiguity which complicate rocket or satellite measurements. In section 2 we describe the event and data used; section 3 outlines the technique and methodology; in section 4 we show two typical examples of PA patches; section 5 shows statistical results; in section 6 we discuss our results, and finally, in section 7 we summarize and draw conclusions.

## 2. Data and Conditions

This study utilizes all-sky imager data obtained at Poker Flat Research Range in Alaska, located at 147.4/65.1° geographic longitude/latitude (~65.5° magnetic latitude). The full-frame 512 by 512 pixels images have a 1000 km field of view (FOV) of the sky at the assumed altitude (110 km) of the emissions. Throughout the recorded movies, the sky is clear, the moon is down, and there are no artifacts such as street lamps. The all-sky imager was filtered for the 557.7 nm green line from atomic oxygen and was operated at 3.3 Hz frame rate, resulting in spatial and temporal resolution that are well suited to study the diffuse pulsating aurora.

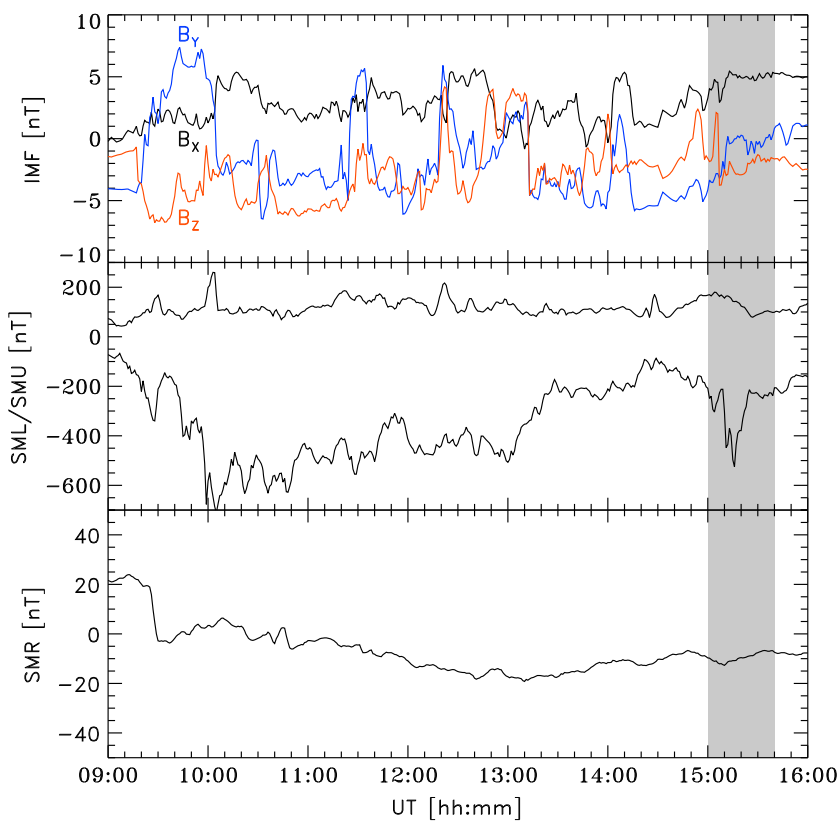

**Figure 1.** (top) IMF, (middle) SML/SMU indices, and (bottom) SMR. The time of the movie is indicated by the grey box. We use the SuperMAG data set of indices and time shifted IMF.

### 2.1. The 1 March 2012 15:00–15:40 UT

Our event occurred on 1 March 2012 during the late expansion and recovery phase of a substorm, as evident from the SMU index in Figure 1 (middle). We use the SuperMAG data set [*Gjerloev*, 2012; *Newell and Gjerloev*, 2011a, 2011b] of indices and propagated interplanetary magnetic field (IMF), which can be obtained through the SuperMAG website. The geomagnetic conditions are quiet to moderately disturbed with a moderately southward interplanetary magnetic field (IMF) and almost no ring current activity, as shown in Figure 1 (top and bottom). However, the SMU index shows 4.5 h of almost continuous substorm activity starting with the onset at 9:41 UT preceding our event. According to the SuperMAG substorm database [*Newell and Gjerloev*, 2011a], a second substorm onset happens at 14:48 UT. Before 15:00 UT the pulsating aurora covers the southern part of the sky with large east-west structures which seem to pulsate in a streaming fashion. At about 15:00 UT the pulsating aurora starts to break up into smaller more distinct and persistent patches. Our selected patches are located postmidnight around 4 magnetic local time (MLT).

## 3. Technique

To determine the characteristics of individual pulsating patches, we perform a five-step analysis:

1. Perform a cartesian projection.
2. Manually identify a pulsating patch.
3. Correct for Earth's rotation.
4. Contour and extract the patch.
5. Determine the patch properties.

These steps require a brief explanation.

### 3.1. Step 1: Cartesian Projection

The fisheye lens of the all-sky imager provides the maximum spatial coverage, but it also produces a strongly distorted view of the night sky. To correct for this, we transform each of the images to a square frame with a uniform pixel resolution of 2.0 by 2.0 km as shown in Figure 2. First, we transform the original spherical geographical pixel coordinates (at 110 km altitude) to Cartesian geographical coordinates (assuming $r = 1$). In this way we can use the scalar product to find the pixel vector in the original image that lies closest in geographical coordinates to each pixel vector in the new image. The resolution is preserved, and we do not need to make any assumptions of the altitude of the emissions. The only artifact is that the emissions in the outer parts of the image are collected over a larger area than covered by the new pixel size and likewise that emissions in the center of the image are collected over a smaller area than covered by the new pixel size. To avoid the most distorted limb pixels of the images, we only include patches located within the center of the FOV (a 500 by 500 km square).

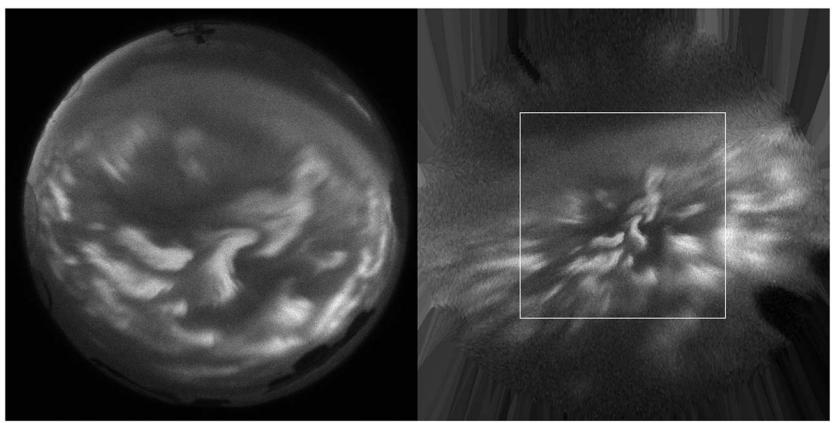

**Figure 2.** (left) Distorted fisheye view ASI image (right) transformed to a uniform 4 km$^2$ pixel grid. Patches outside the white frame are not analyzed to avoid the most distorted limb pixels.

### 3.2. Step 2: Patch Identification
Individual patches are manually identified from the movie and a keogram (see Figure 3). The keogram is merely used to decide if the patch is pulsating and for how long we can follow it before it either disappears or, for example, joins together with an adjacent patch.

### 3.3. Step 3: Correct for Earth's Rotation
The apparent drift of a patch in the movie frame of reference can be due to a combination of the Earth rotation (moving the FOV across the night sky) and the drift of the patch in an inertial frame. This latter frame is effectively GSE coordinates. We determine a rough contour around the patch for its entire lifetime to determine the apparent drift of the patch. A new movie of each individual patch is made where the patch is centered at all times (see Figure 4). This technique assumes that the velocity is constant during the lifetime of the patch and none of the patches appear to violate this assumption. The Earth's rotation is known and is easily subtracted from the apparent drift velocity to allow for a determination of the patch inertial frame velocity.

### 3.4. Step 4: Contour and Extract the Patch
The extracted patch is expected to vary during the train of pulsations. A new contour of the patch is determined for each pulsation. First, we identify an intensity level $L$ where the patch intensity falls off; $L = (I_{max} - I_{min}) \times P$, where $I_{max}$ is the maximum intensity during the pulsation, $I_{min}$ is the minimum background intensity, and $P$ is the percentage value which best captures the patch contour, for example, to avoid capturing an adjacent patch. For each pulsation the contour is defined as the outermost edge of the sum of contours found for the ±2 images (±0.6 s) around maximum intensity as illustrated in Figure 5. For the times between pulsations the contours are interpolated using a square function.

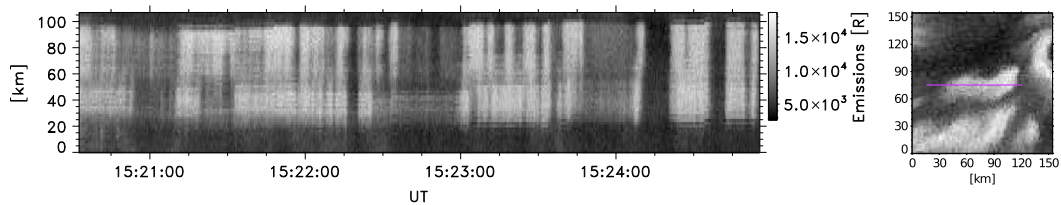

**Figure 3.** Geographic east-west keogram of the resulting time interval for a specific patch (#2). In Step 2 of the technique a keogram through a patch is used to decide if the patch is pulsating and for how long we can follow it.

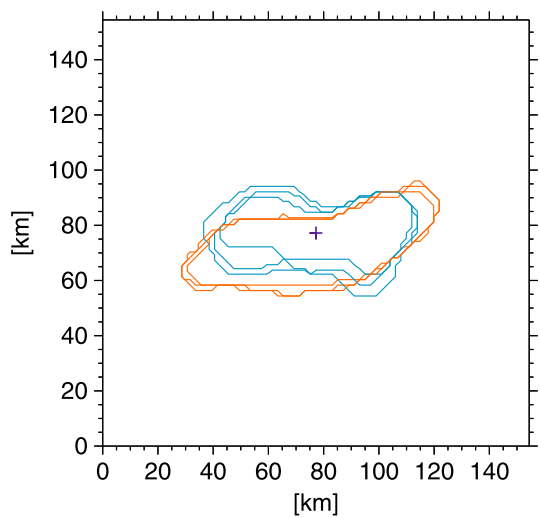

**Figure 4.** Contours of a patch (#2) for the three first (blue) and last (orange) on-times of a patch. In Step 3 of the technique the net velocity of the rotation of Earth and assumed constant drift is corrected for, resulting in centered contours.

In this study we only discuss individual patches. However, it happens that patches merge with one or several nearby patches. Exactly how frequently under which conditions and why this happens are not known. If the patch merges with a neighboring patch, we end the analysis. If a contour cannot be determined because of weak emissions, we use the average of the previous and following contours. To avoid capturing adjacent patches and background emissions, we visually check each contour.

### 3.5. Step 5: Patch Properties

With the patch identified and contoured, we can objectively determine the temporal characteristics and the energy deposition of the patches. For each image we calculate the patch total intensity (units of Rayleigh) as well as the patch median intensity (units of Rayleigh per square kilometer). The properties are calculated from the median intensity because it is less affected by the contouring technique and thus less sensitive to the background emissions.

First, possible pulsations are identified by finding the maxima and minima of the median intensity (we use a 1.5 s boxcar filter to remove local minima). An actual pulsation is defined as having a buildup or decay larger than the pulse threshold. The pulse threshold is 5% of the difference between the maximum and minimum median intensity in the time series. Second, the off-time is the time between pulsations defined as the interval around the minima where the intensity does not vary more than the pulse threshold. The on-time for an individual pulsation is further defined as the time between off-times. Figure 6 shows an example interval of a patch median intensity where the terms on-time and off-time are indicated. Figure 7 (top) also shows the smoothed median intensity (black curve) and the resulting maxima and off-time intervals for a train of pulsations.

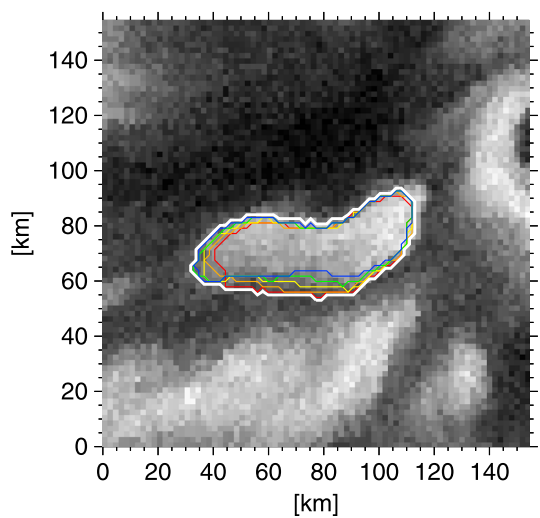

**Figure 5.** For each on-time the contour of the patch is found defined as the outermost edge (white) of the contours (red, orange, yellow, and green blue) found for the maximum intensity image ±2 images, as described in Step 4 of the technique.

To determine the intensity of an individual pulse, we must subtract the background emissions or offset. This may be due to diffuse aurora at higher altitudes, or in other words, we assume that a pulsating patch is superposed onto a background. The offset is estimated from the off-time values, and a linear fit is used to find the offset during the on-time (see Figures 6 and 7 (dashed lines)).

The energy deposition of the patch is referred to as intensity, defined as the median intensity corrected for the offset. Figure 7 (bottom) shows an example of the resulting intensity and train of pulses (highlighted in grey). The resulting off-times are indicated with shaded green areas in the background. In relation to the temporal characteristics, the integrated intensity is the energy deposited over one pulsation/on-time, while the maximum intensity is the peak energy deposition.

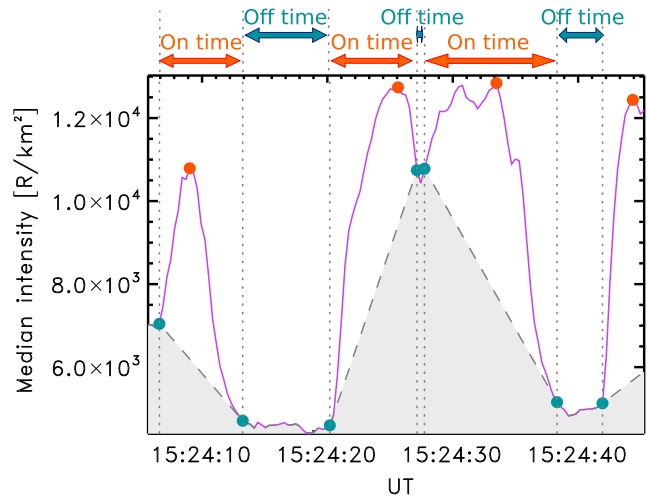

**Figure 6.** An example interval of a patch (#2) median intensity with examples of the defined on-time, off-time, and offset (dashed line).

Using this technique, we identified, extracted, and quantified six patches.

## 4. Typical Example

This section is intended to give an example of what a pulsating patch is as well as to show some typical events supporting the statistical results (section 5).

To illustrate our technique and the behavior of a typical patch (#2), we show eight images from a ~4 min interval (Figure 8). Figure 8 (middle) shows its train of pulsations both in median and total intensity. The images are alternating peak and off-times, and it is apparent that during the off-time the patch still has emissions slightly higher than the surrounding background. We follow the patch for about 250 s or 26 pulses after which it merges with an adjacent patch and we terminate our analysis. It is evident from the median as well as the total intensity that the intensity during the on-times varies from pulse to pulse. The pulsating behavior is highly complex with pulsation on-times ranging from about 3 to 18 s. The patch certainly falls within the broad definition of "pulsating aurora," but the variable period does not appear to be in agreement with the term "pulsating" which by definition is a regular periodic behavior. A more appropriate term may be *fluctuating* aurora which by definition is irregular. Figure 9 shows another typical patch (#4) in the same format as Figure 8. In agreement with the previous example we find a striking variability in basically all measurable parameters. The patch shape may be the only parameter that remains relatively constant throughout the lifetime of the patch (in itself a remarkable observational fact).

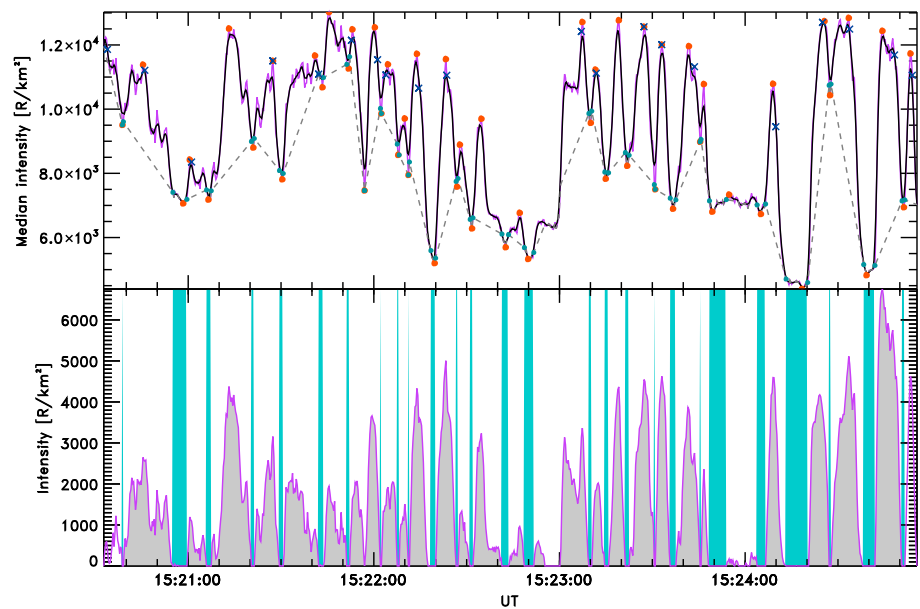

**Figure 7.** (top) The median intensity (purple), smoothed median intensity (black), the resulting maxima (orange) and the corrected minima (green) which define the off-times and on-times of the pulses, and the offset (dashed line). The crosses (dark blue) are the times for which the contours were found as described in Step 4 of the technique. (bottom) The resulting pulses (highlighted in grey) defined as the median intensity corrected for the offset. The off-times are indicated with shaded green areas in the background.

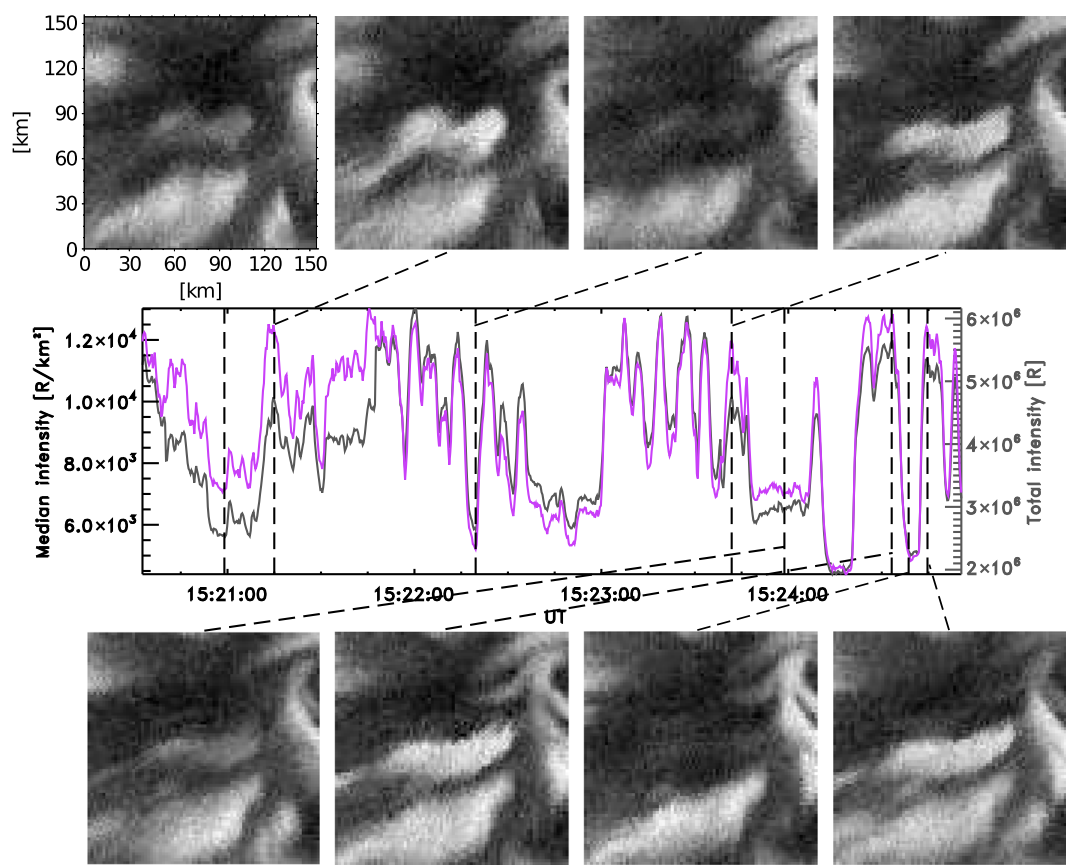

**Figure 8.** The train of pulsations for a typical patch (#2). Median (purple) and total intensity (grey) within the patch and eight image samples where the patch is either considered on/bright or off/dim.

## 5. Statistical Results

With the above definitions we can now determine the temporal characteristics and the characteristics of the energy deposition from each of the six patches.

### 5.1. Temporal Characteristics

As mentioned, Figure 8 displayed a striking temporal variability. Using all six patches, we show in Figure 10 the PDF (probability density function) of the on-times. The distribution shows a large spread of on-times, which vary from 2 to 21 s, with a peak at about 4 s. This reflects the highly variable total and median intensity through the lifetime of a patch, as, for example, shown in Figure 8 of example patch #2. The average on-time is $5.67 \pm 0.14$ s, and the average off-time is $0.80 \pm 0.04$ s. The average on-time and off-time combined form an approximate average period of $6.5 \pm 0.2$ s. Although the spread is wide, the distribution is clearly not scaleless and thus indicates that there is a preferred (typical) on-time of $\sim$3–5 s.

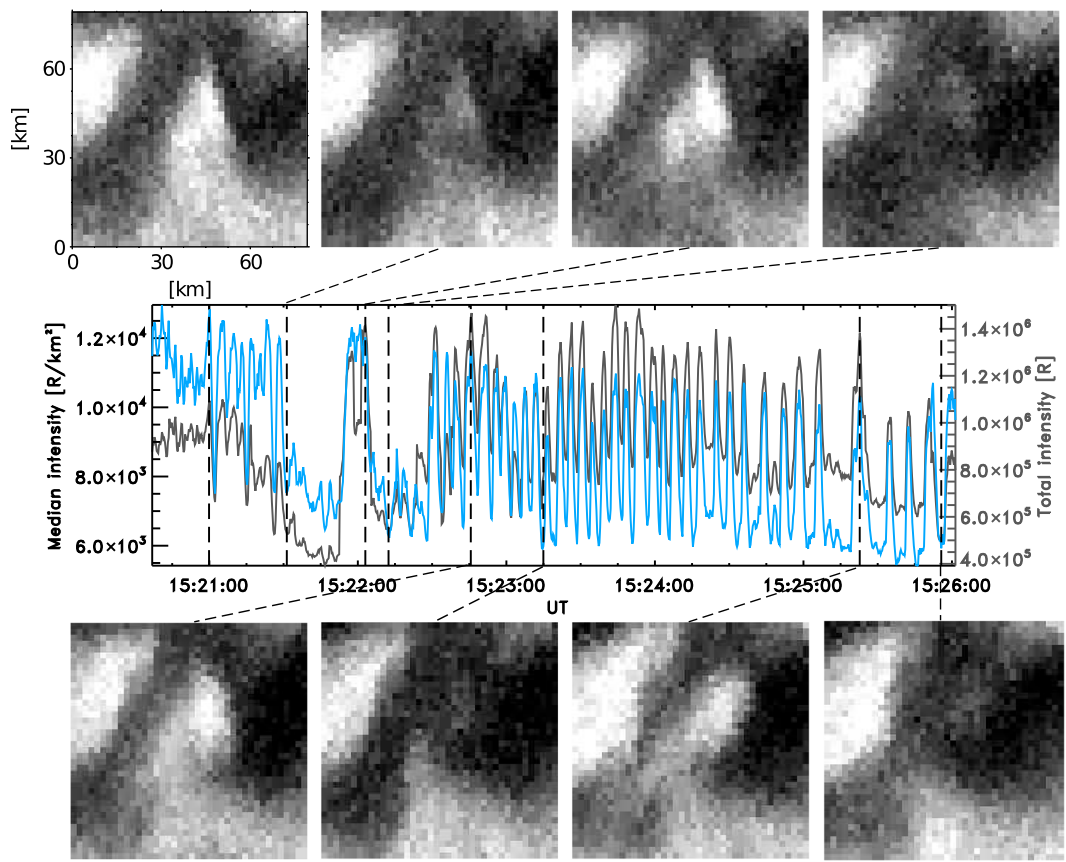

**Figure 9.** The train of pulsations for a typical patch (#4). Median (blue) and total intensity (grey) within the patch and eight image samples where the patch is either considered on/bright or off/dim.

Table 1 lists the duration (time and number of pulses) of the analysis for each of the patches. The exact lifetimes of the patches are probably longer because we start to capture most of the patches a bit into their lifetime and our technique is sensitive for neighboring patches which might terminate the analysis before the patch has disappeared. It is, however, clear that most of the patches are incredibly persistent.

### 5.2. Energy Deposition

We estimate the energy deposition for each pulse by integrating the intensity over the on-time interval. For this we use the median intensity rather than an actual 2-D spatial integration since this is less sensitive to uncertainties in the contouring procedure. This simplification can be described as $E_{\text{tot}} = \sum_{i=1}^{n} \sum_{j=1}^{512} \sum_{k=1}^{512} I_{i,j,k} A_{\text{pixel}} \approx \sum_{i=1}^{n} \bar{I}_i A_{\text{patch}}$ where the summation is over time or frame ($i$), the $x$ coordinate ($j$), and the $y$ coordinate ($k$), $A_{\text{pixel}}$ is the pixel area, $A_{\text{patch}}$ is the patch area, and finally, $\bar{I}_i$ is the median of the total intensity within the patch.

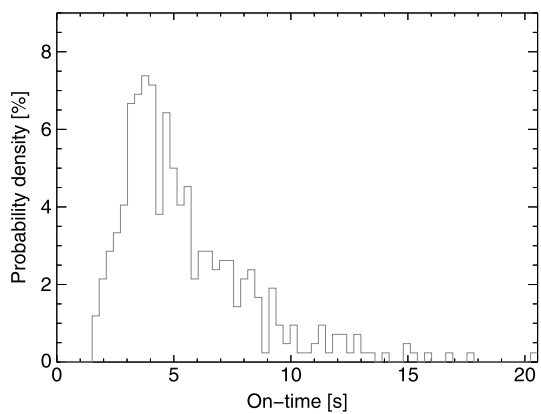

**Figure 10.** On-time probability density of the pulsating auroral patches.

The result, of course, does not have units of energy although the intensity of the green line is roughly proportional to the total energy flux of precipitating electrons. Further, this analysis does not include Poynting flux. We should, thus, be careful of interpreting it as a measure of energy deposition, but for the purpose of estimating

Table 1. The Duration (Time and Number of Pulses) of the Analysis for Each of the Patches

| Patch No. | Time (min:s) | No. of Pulses |
|---|---|---|
| 1 | 12:05 | 127 |
| 2 | 4:12 | 26 |
| 3 | 14:35 | 132 |
| 4 | 5:16 | 54 |
| 5 | 4:23 | 40 |
| 6 | 5:05 | 36 |

the precipitating electron energy deposition, this technique should produce representative results.

Figure 11 shows the integrated intensity for all patches and all on-times. The scatterplot indicates a weak correlation ($r = 0.7$), but this is to be expected since the duration of the on-time ($X$ axis) is obviously linearly correlated with the integrated intensity ($Y$ axis). The simple linear fit suggests that the energy dissipated can be approximated by $f(T_{on}) = (5800 \pm 300)T_{on} - (1900 \pm 1800)$ R/km$^2$. However, the linear relationship is purely extrapolation as significant scatter is present, especially for longer on-times (exceeding ∼6 s).

We could hypothesize that for the lifetime of a patch each pulsation (on-time) would involve the same amount of energy deposited. That would imply that a longer on-time would be associated with a weaker intensity. Figure 12 shows the maximum intensity as a function of on-time. What we find is virtually no correlation ($r = 0.27$) between the on-time and maximum intensity. Although the significance of the relationship is so that we can reject the null hypothesis, the very weak relationship implies that for the lifetime of the patch the energy deposition of the patch varies from one pulsation to another. It is also revealed that the maximum intensity varies through a train of pulsations, implying that the PA patches do not have a preferred maximum intensity.

Further, we could hypothesize that the off-time (wait time) is related to the subsequent energy deposition. The idea here is a storage and release which would then lead to a relationship where longer off-times lead to larger energy deposition. In Figure 13 we show the energy deposition as a function of the preceding off-time. The steps on the $x$ axis represent the time resolution of the ASI data. A longer off-time does not imply a larger amount of energy deposited over the next on-time.

Finally, we test how the energy deposited changes during the on-times. Figure 14 shows the median of the on-times normalized in time and intensity. We separate the on-times into "long" duration (on-time >6 s, green line) and "short" duration (on-time ≤6 s, pink line). The latter captures the bulk of on-times (see Figure 10). This is done in an attempt to decrease the scattering due to double peaks and multipeaks, which likely are more frequent the longer the on-time is. However, also relatively short on-times have these characteristics (see, for example, second pulse at about 15:21 UT in Figure 7, bottom). In general, the energy deposition is highly irregular which results in considerable scatter for both short- and long-duration on-times. There is a tendency for short on-times (pink) to have a more well-defined peak, reaching an intensity level of about 0.95, compared to the long on-times (green) that stabilizes at an intensity level slightly less than 0.8 for roughly 35% of the on-time. The short-duration energy deposition (pink) is remarkably symmetric. The long-duration energy deposition (green) is symmetric below a normalized intensity level of about 0.55, but if we look at the entire on-time, the energy deposition has a quicker buildup than decay. The asymmetric distribution could be interpreted as indicative of a rapid release and slower decay in support of a loading-unloading process.

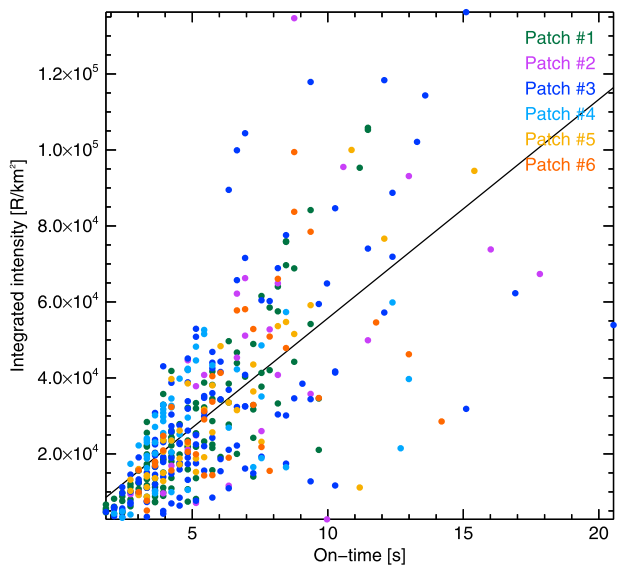

**Figure 11.** The intensity integrated over the on-time compared to the on-time. The different colors represent the different patches.

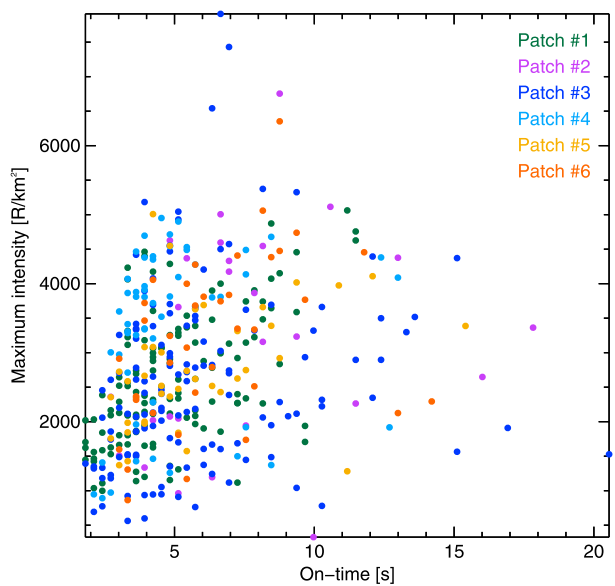

**Figure 12.** Maximum intensity during the on-time compared to the on-time. The different colors represent the different patches.

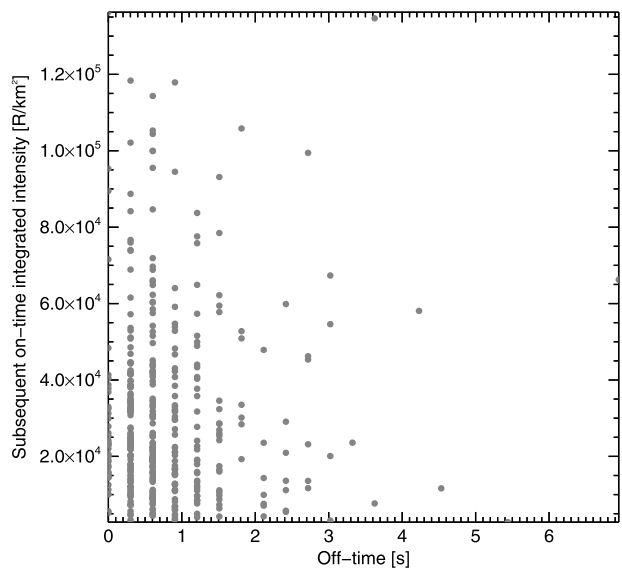

**Figure 13.** Off-time compared to the intensity integrated over the subsequent on-time.

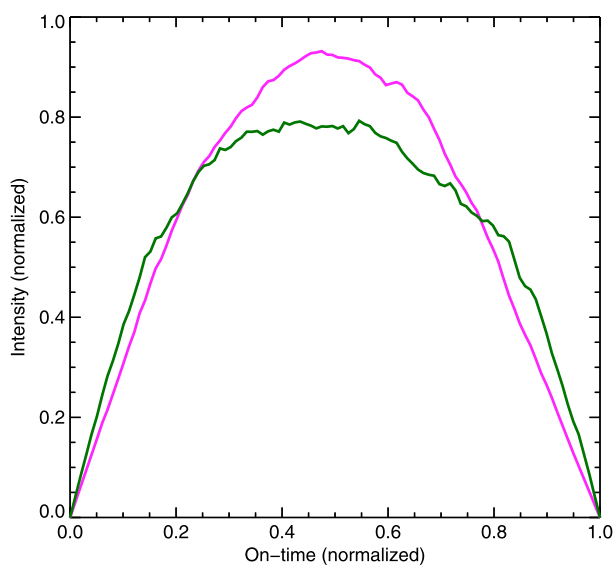

**Figure 14.** The median energy deposition of all pulses normalized in time and intensity. The on-time is separated into long on-times greater than 6 s (green) and short on-times less or equal to 6 s (pink).

## 6. Discussion

To interpret our results, we must first address the inherent limitations of the ASI data set and technique. The main limitations are as follows: (1) observations are limited to six patches, (2) 3.31 Hz time resolution, (3) 557.7 nm emissions, and (4) finding the exact area of the patches. The resulting characteristics and observational constraints of PA patches on the suggested mechanism will then be discussed. Finally, we discuss the majority of the most frequently mentioned theories of pulsating aurora that are listed in Table 2.

### 6.1. Inherent Limitations

The six patches analyzed are carefully selected adjacent patches. The amount of data available is large, but the technique chosen for this study is time consuming. The limited observations (1) of course result in limited statistics, and we cannot address how the characteristics depend on geomagnetic conditions, especially substorm phases, local time, latitude, or any other parameter that may vary from event to event. The pulsating auroras are observed under a wide variety of conditions [e.g., *Jones et al.*, 2013], meaning that this study may only address a small subset. Acknowledging this opens up another question as to the external or solar wind control of the patch evolution and behavior, a topic well beyond the scope of this paper. The limited observations could therefore be argued to be a distinct advantage since all occur during the same event and thus any variability cannot be due to dependencies such as geomagnetic activity, local time, and so on.

The 3.31 Hz time resolution (2) of the ASI data used is limited by the Nyquist frequency of 1.65 Hz. Therefore, we cannot distinguish rapid temporal behavior such as the ∼3 Hz modulations often found to be superimposed on the slower variations and so-called flashes. Flashes are burst-like fluctuations where the on-time is shorter than 1 s, while the off-time can reach several minutes [*Tsuruda et al.*, 1981; *Yamamoto*, 1988]. Running the all-sky cameras at a higher cadence is possible, but the pulsating aurora that occurs with higher-frequency fluctuations, such as the 3 Hz modulations, occurs at small spatial scales on top of the large-scale pulsations. Therefore, to fully investigate the properties of the higher frequencies, a narrow field of view imager is more appropriate as was done in *Samara and Michell* [2010], where pulsations up to 10 to 15 Hz were reported. It is entirely possible that these higher-frequency pulsations are caused by completely different mechanisms, and therefore, it would make sense to investigate them independently.

The 557.7 nm emissions (3) are limited by a chemical effect. Compared to prompt emissions, the 557.7 nm emissions are found to have a mean lifetime of 0.3 to 0.59 s in pulsating aurora, the highest values of mean lifetime found in observations of sharp-edged patches *Scourfield et al.* [1971]. The important result is then a temporal smoothing over the same time scale, meaning that the 557.7 nm filter used could smooth out possible impulsive behavior. This study therefore focuses on the main on-off (bright/dim) modulation of the PA patches.

Table 2. The Most Frequently Mentioned Theories of Pulsating Aurora, Their Auroral Observational Consequences, and How They Fit Within the Observational Constraints of Pulsating Auroral Patches

| Theory | Summary | Auroral Observational Consequences | Can It Explain PA Patches? |
|---|---|---|---|
| Nonlinear relaxation oscillator [*Davidson*, 1979, 1986a, 1986b] | The control of wave growth is attributed to changes in the anisotropy as the loss cone is filled and emptied. The particles that are lost need to be replenished for the cycle to start again. | – Periods of 3 to 30 s<br>– Rapid rise in energy deposition corresponding to wave growth times of ~0.1–1 s.<br>– The most probable period is near 10 s. | Yes |
| Nonlinear pulsation model [*Davidson and Chiu*, 1991] | Temporal variations may be driven by spatial variations in plasma density through a nonlinear pulsation model. | – Irregular pulsations<br>– Trains of pulsations that begin between the pulses of a train that started several cycles earlier<br>– Fourier spectrum of mainly broad features | Yes |
| Flow cyclotron maser [*Demekhov and Trakhtengerts*, 1994] | A continuous stream of energetic electrons enters a flux tube with enhanced cold plasma density that serves as a resonance cavity. Low-frequency waves start to scatter the high-energy electrons into the loss cone. The scattering moves toward higher wave frequencies and lower energy electrons and thus includes more particles. This continues until a maximum of the electrons participates and the waves are damped. | – On-time is determined by the nonlinear dynamics and is on the order of wave travel time between the ionospheres ~1–5 s.<br>– Off-time is nearly equal to the time of energetic electron accumulation in the duct, suggesting a repetition period of ~5–30 s.<br>– Slower buildup than decay of the energy deposition | No |
| Backward wave oscillator [*Trakhtengerts*, 1995, 1999] | Waves propagate opposite to the motion of interacting electrons resulting in step-like interactions and the generation of chorus bursts composited of chorus elements. | Rapid fluctuations superimposed on the on-time of quasiperiodic variations | Unknown |
| Flow cyclotron maser with ionospheric feedback [*Tagirov et al.*, 1999] | The feedback is due to the local increase in ionospheric electron density that reduces the waves and thus the scattering process at magnetic equator until it stops. This effect lags several seconds to the ionization by precipitating particles. As the electron density decreases due to recombination, it allows waves to begin to grow once again. | – Relaxation time is about twice the buildup time.<br>– Whole cycle of the order 5–15 s<br>– PA patch lifetime of 1–3 min | No |
| ULF waves [*Coroniti and Kennel*, 1970] | ULF waves act to modulate the velocity distribution of resonant electrons and thus the pitch angle scattering at magnetic equator. | – Relative periodic scattering dependent on diffusion rate and ULF wave frequency<br>– Modulation should be superposed on an already enhanced precipitation background. | Unknown |

**Table 2.** (continued)

| Theory | Summary | Auroral Observational Consequences | Can It Explain PA Patches? |
|---|---|---|---|
| Fermi-type acceleration [*Nakajima et al.*, 2012] | Field-aligned electrons are already generated by Fermi-type accelerations associated with earthward plasma flow at substorm onset. Pulsations of the field-aligned electrons are further suggested to be created by small pitch angle modulations by weak whistler mode waves, ECH waves, and/or magnetosonic waves, or alternatively Alfvén waves. | | Unknown |
| Ionospheric ion-acoustic instability [*Pilipenko et al.*, 1999] | Ion-acoustic instability can arise when densities of local field-aligned currents in the topside ionosphere reach the threshold for excitation of high-frequency turbulence. Further, the ion-acoustic instability gives rise to quasi-oscillatory variations of parallel electric fields and consequently particle acceleration and precipitation. | Quasi-oscillatory particle precipitation | Unknown |
| Auroral acceleration region modulation by Alfvén waves [*Fedorov et al.*, 2004] | Magnetospheric Alfvén waves penetrating into the auroral acceleration region can produce oscillatory variations of the field-aligned potential drop in the auroral topside ionosphere | May cause oscillatory frequency dependence of electron acceleration modulations in the range around fractions of a hertz. | Unknown |
| Atmospheric pressure waves [*Luhmann*, 1979] | Pressure waves in the neutral atmosphere cause quasiperiodic fluctuations in the auroral intensity by changes in the altitude at which particles are lost to the atmosphere and thus the size of the loss cone. | – Quasiperiodic fluctuations in auroral intensity dependent on the frequency of the atmospheric density fluctuations and magnitude and direction of the local neutral wind<br>– Can only explain PA periods longer than the bounce time | Unknown |

The main technique limitation is finding the exact area of the patches (4) using the contouring technique as outlined in Step 4. For each image we calculate the patch total intensity (units of Rayleigh) as well as the patch median intensity (units of Rayleigh per square kilometer). To limit the uncertainty in finding the exact area of the patches, we use the median intensity which is less affected by the contouring technique. It could also be discussed whether it is right to use the whole patch area compared to, for example, defining a core. The temporal variation we show is a smoothing over the scale size of the patch, which effectively smooth out any spatiotemporal variability within the patch and can make it difficult to put a threshold on what is considered sufficient rise and decay in intensity used to define the on-time and off-time of the patch. Then again, it is equally or more difficult to define, for example, a core area.

### 6.2. What Is the Temporal Variability of PA Patches?

We have defined an on-time and off-time of the patches and do not use the common term period because we find the temporal variation of the brightness of pulsating aurora to be highly variable and far from periodic or even quasiperiodic. Our technique has the advantage that it objectively determines the intensity variation using the patch area. It also provides a clear definition of the on-time and off-time and removes the unwanted variations that are due to the Earth's rotation and drift of the patches. In comparison, other published studies used simple north-south keogram or a sampling box around a central portion of the patch.

In the literature typical pulsating periods are referred to range from 2 to 20 s [*Royrvik and Davis*, 1977]. The results shown in Figure 10 show on-times in the range of 2 to 21 s, which is in good agreement with other published studies. We do, however, find more of the shorter than longer on-times. The average on-time ($5.67 \pm 0.14$ s) and off-time ($0.80 \pm 0.04$ s) combined form an approximate average period of $6.5 \pm 0.2$ s, which is shorter than the typical average period of $8 \pm 2$ s found by *Royrvik and Davis* [1977]. It should be mentioned here that our skewed PDFs indicate that the use of average is not appropriate. Rather, we should use typical (or mode), which may be more appropriate in describing the behavior. This means either that the pulsating auroral patches have a different characteristic compared to other kinds of pulsating aurora or that the use of the term period is simply not a characteristic for the pulsating aurora.

The large difference in on-times and off-times also suggests that the use of the terms on-time and off-time fits the fundamental characteristics of pulsating aurora better than the frequently used terms period, recurrence period, and on-off period. A similar conclusion was drawn by *Yamamoto* [1988] based on findings of mostly individual isolated pulses where the pulsation on-time therefore became the most essential quantity. He found that the standing and streaming forms combined have a median on-time of 3–4 s and a median off-time of 4.5–7 s for the *Kp* indices 3–4. The median of the 420 on-times shown in Figure 10 is 4.8 s, and the median off-time is 0.6 s (based on 426 off-times, not shown). The median on-time is thus slightly longer, while the median off-time is much shorter and the variance in the data set is larger for the on-times than the off-times, opposite to what was found by *Yamamoto* [1988]. He defined the on-time as the width of the pulse and off-time as the separation of individual pulsations, similar to what we do, but used a correlation to find the times. There are, however, factors that can explain the difference. The most important one is that the so-called type 1 pulsations are included in their statistics, while the type 3 pulsations might not be included. The type 1 is classified as a series of pulses having shorter on-times than off-times, type 2 as quasiperiodic fade-outs or decreases in luminosity from the on level having longer on-times than off-times, while the type 3 pulsations show neither full luminosity changes nor a clear distinction of on-times and off-times. When simply comparing the typical example in Figure 8 to the example Figures 1–3 and 5 in the paper by *Yamamoto* [1988] showing the different types of pulsations, the best fit is found to be either type 2 or type 3, consistent with a typically longer on-time than off-time. The different median on-times and off-times reported can therefore result from our observations of the type 2 and type 3 pulsations and not the type 1 pulsations. It should, however, be noted that the pulsating aurora analyzed in this study is relatively well-separated patches of either standing or streaming mode, while *Yamamoto* [1988] also included the moving mode, which is far more difficult to objectively identify. This example also illustrates how complicated it is to find the temporal characteristics, and not to mention a classification, of the pulsating aurora.

Although the spread is wide, the distribution is clearly not scaleless and thus indicates that there is a preferred (typical) on-time of ~3–5 s. Most later studies such as *Nishiyama et al.* [2014] and *Sato et al.* [2015] avoided the difficulties of defining the on-time and off-times by using the recurrence period and on-off period, respectively, as temporal characteristics of the main pulsation. *Sato et al.* [2015] found recurrent periods of ~9–12 s estimated from auroral intensities along a north-south keogram, but as they clearly point out, they investigate the characteristics of omega band pulsating aurora, which is significantly different in their shapes and structure compared to PA patches and possibly have a different generation and modulation mechanism. *Nishiyama et al.* [2014] studied the dual-scale temporal characteristics of PAs, meaning the relationship between the on-off periods and rapid modulations. The on-off periods, which were estimated with a fast Fourier transform analysis with a 60 s data window, ranged from 1.6 to 15 s. PA with on-off periods shorter than 10 s accounted for 72% of the total events in their statistical analysis, and as can be seen from their Figure 4, they find most on-off periods from 4.0 to 8.0 s, which is consistent with our result and thus supports the suggestion of a preferred (typical) on-time of ∼3–5 s. Both the preferred on-time and short off-time are characteristics that should be explained by the suggested mechanism(s) of pulsating auroral patches.

### 6.3. PA Patches Energy Deposition in Relation to Their Temporal Characteristics

This is the first study attempting to estimate the energy deposition in association with the on-time and off-time of PA patches. We have done an analysis of the energy deposition, where the auroral intensity is used as an indicator of the actual energy deposition by the energy flux of downward electrons. Although we may argue that the emissions are related to the particle precipitation, they do not include Poynting flux. However, a comparison between precipitating energy flux and the relative intensity of the auroral luminosity extracted from a corresponding all-sky image was found to agree well [*Stenbaek-Nielsen et al.*, 1998]. The analysis of estimating the energy deposition from auroral emissions is out of scope of this study. A few simple hypotheses relevant for the suggested mechanisms and drivers of the pulsating aurora are tested. We find the following: (1) A weak correlation between the length of the on-time and the energy deposition. This is as expected, but there is significant scatter, especially for longer on-times. (2) No correlation between the on-time and maximum intensity. (3) PA patches have no preferred maximum energy deposition. (4) The off-time does not have a relation to the amount of energy that is deposited over the next on-time. (5) Evidence of a slightly quicker build up than decay of the energy deposition.

According to our observations, the PA patches have no preferred maximum energy deposition. However, it has been reported that the intensities of the PA patches have relatively constant maximum amplitude. The difference might be due to the offset subtracted, which has a slower variation. For example, *Yamamoto* [1988] reported that the amplitude was fairly constant throughout the train of pulsations and noted that the maximum intensity showed much slower changes than the period of pulsation, indicating that the background conditions, which may relate to the causal instability, vary little during several cycles of pulsations. The train of pulsations in Figure 8 shows evidence of the same, where the on-off pulsations seem to be superposed on a slower variation in the background intensity. We have taken this into account and subtracted the offset before investigating the energy deposition (see Figure 7). A further investigation of the background emissions are out of scope of this paper, but the observation can be of interest in the ongoing research on the role of the background plasma as driver of the chorus emissions. For example, *Li et al.* [2011a] did find a one-to-one variation in the ULF and chorus modulations with periods ranging from tens of seconds to a few minutes, and *Jaynes et al.* [2015] link chorus modulations with ∼45 s to 1 min periods to ULF waves having periods closer to 2 min. Interestingly, the train of pulsations of patch #2 in Figures 7 and 8 indicates that the offset has a rise and decay over about one to four on-times. In addition, purely visually, it looks like there also exists a slower background variability of about 70 s (15:21:00–15:22:20 UT), 30 s (15:22:20–15:22:50 UT), and 90 s (15:22:50–15:24:20 UT). Also, the train of pulsations of example patch #4 in Figure 9 gives an indication of this dual-temporal characteristic of the background emission. In contrast to the results by *Yamamoto* [1988], we find that the maximum intensity varies through a train of pulsations, revealing that the PA patches do not have a preferred maximum intensity and that there probably are complex varying precipitation fluxes behind the on-off pulsations. The different results likely arise because we subtract the offset to study the characteristics of the on-off pulsations, while the maximum amplitude by *Yamamoto* [1988] included the slower varying background emissions.

Our results in Figure 12 show that there is virtually no correlation between the on-time and maximum intensity. One might simplistically expect a relationship if the total energy deposited by the patch was approximately constant from pulse to pulse. It has been suggested that the on-off pulsations are closely correlated to the amplitudes of lower band chorus waves [*Nishimura et al.*, 2010, 2011] or electron cyclotron harmonics [*Liang et al.*, 2010] at the magnetic equator. The amplitude of the waves at magnetic equator tells us how effective the pitch angle scattering rate is. This implies that the available energy simply would be deposited over a shorter on-time (negatively correlated), which is not the case. Interestingly, this seems to disagree with a recent study by *Samara and Michell* [2010]. They found that the frequency of the pulsations correlates to the intensity of the aurora, with the brighter aurora being associated with the higher frequencies. The disagreement between the observations is likely because the higher frequencies correlate to the intensity and not the on-time. This is confirmed by a later observation of the auroral intensity being correlated to the ~3 Hz frequencies and not the on-off periods [*Nishiyama et al.*, 2014]. The observations combined support the proposed explanation that the higher frequencies superposed on the on-time intensity are related to the chorus elements and that closer spacing between the chorus elements corresponds to higher-energy resonant electrons [*Nunn et al.*, 2009; *Samara and Michell*, 2010]. Our finding suggests that the total energy deposited by the patch varies from pulse to pulse. The variable energy deposition can possibly be due to the on-off pulse being a composition of higher-frequency pulsations, where the composition for some reason varies from pulse to pulse. We find no preferred relationship between maximum intensity and the on-time in support of the auroral intensity being correlated to the superimposed higher frequencies and not the frequencies in the on-off regime.

Figure 14 indicates a slightly quicker buildup than decay of the energy deposition for the longer on-times (>6 s). This is the opposite of the results by *Samara and Michell* [2010] that also found the intensity profile of the pulsating structures to be asymmetric, but with the intensity increasing at a slower rate than it decreases. They suggest that this could be consistent with the backward wave oscillator chorus generation mechanism [*Trakhtengerts*, 1995, 1999], where the chorus bursts increase in frequency until they reach a cutoff and no longer resonate with the plasma sheet electrons of the correct energy range (few keV to tens of keV). They find it to be a common feature in the many different pulsating auroral structures examined but most pronounced during a time period where our observations are not entirely comparable. The main differences being the magnetic local time (~0400 MLT versus ~0030 MLT), the time from a substorm onset (1.5 versus 0.5 h), and their observation of several distinct pulsating structures and pulsations forming at the southern edge of a brighter auroral arc, but not well-separated pulsating patches. It is, however, unclear how these differences would explain or influence the generation mechanisms. Further, the above general picture of chorus generation can be verified utilizing pulsating aurora, especially the pulsating patches as manifested in the flow cyclotron maser theory [*Trakhtengerts*, 1999]. Hence, we find it puzzling that we do not observe evidence of the bursts of several chorus elements which are suggested to appear in the final stage of a so-called optical flash. The indication of the energy deposition having a slightly quicker buildup than decay for the longer on-times (>6 s) and a symmetric buildup and decay for the shorter on-times (≤6 s) might therefore not support the observational constraints set by the flow cyclotron maser theory, which is suggested to create the pulsating auroral patches.

The slightly quicker buildup than decay of the longer on-times (> 6 s, green line) in Figure 14 agrees with the observations by *Jaynes et al*. [2013]. They often found a sawtooth component in their observation of the electron precipitation close to the magnetic equator. The same feature was also evident in the luminosity of the pulsating auroral patches that correlated to the particle measurements. Their observations were interpreted as a possible loading and dumping cycle for the particle populations that fit well within the framework of the nonlinear relaxation oscillator theory. They sketch a scenario where the diffusion becomes less and less effective after the initial filling and emptying of the loss cone, possibly multiple times, as the electrons at adjacent pitch angles continue to diffuse into the loss cone during multiple bounces through the near-equatorial interaction region. The duration of the sawtooth with a rapid rise of 0–4 s and more gradual decrease of about 20 s does not fit to the on-times that we detect (6–21 s where most are less than 10 s and only one above 20 s), and we do not see a clear sawtooth in Figure 14. However, the indications of a slightly quicker buildup than decay of the energy deposition for the longer on-times (>6 s) can still be indicative of a rapid release and slower decay in support of the nonlinear relaxation oscillator theory.

When comparing auroral emissions to the suggested mechanisms and drivers of PA, there are a few effects we should keep in mind. As already mentioned, we observe the 557.7 nm emissions, which introduce a time lag (<0.6 s) and more importantly a smoothing over the same time scale, and we perform a quantitative analysis using the auroral intensity as indicative of the energy deposition. Additionally, there is the effect from time dispersion. In a beam of energetic electron precipitation the highest-energy electrons reach the atmosphere first and will at ionospheric altitudes result in an energy dispersion. This has been used to trace the source region of the pulsations assuming that the pulsations are imposed at one point. A simple exercise using the experimental measurements of bounce times above Poker Flat during moderate disturbed times [*Nemzek et al.*, 1992] shows that the lag of 10 keV to 30 keV electrons traveling from the magnetic equator could be in the order of ~0.4 s. In situ measurements by sounding rockets detected pulsating precipitation of electrons with energies from a few keV to tens of keV [*Bryant et al.*, 1975; *Sandahl et al.*, 1980; *Yau et al.*, 1981]. *Bryant et al.* [1975] observed pulsations in energies from 3 keV and up, the most pronounced pulsations >9 keV, while *Sandahl et al.* [1980] found clear pulsations in electron flux from 5 to 40 keV, the most pronounced around 20 keV. We also calculated the length of a magnetic field line from Poker Flat to magnetic equator using the Tsyganenko 2001 (T01) field model [*Tsyganenko*, 2002a, 2002b] and found that the electron precipitation times were comparable to the results by *Nemzek et al.* [1992]. The assumption here is of course that no field-aligned potential drops are located between the plasma sheet and the observed ionospheric emissions. The travel times we calculated were 1.5 s for 3 keV, 0.9 s for 10 keV, 0.5 s for 30 keV, 0.4 s for 50 keV, and 0.3 s for 100 keV precipitating electrons traveling 7.8 $R_E$. This means that the time lag between the 3 keV and the 100 keV pulsation could reach a considerable 1.2 s. However, the most probable time lags are 0.6 s or less if we assume, according to the in situ measurements, that the pulsations mostly occur above 10 keV. The dispersion effect is further complicated if we account for details of the suggested pitch angle scattering mechanism, for which the wave packet originates at the magnetic equator and propagates toward higher latitudes, resonating with electrons of increasing energy traveling the opposite way [*Miyoshi et al.*, 2010]. Even though the resonance of higher-energy electrons occurs after that of lower energy electrons and the higher-energy electrons travel a longer path, *Miyoshi et al.* [2010] found that the scattered electrons at higher energies overtake the electrons at lower energies before reaching the atmosphere. Taking into account this effect therefore acts to decrease the time dispersion. The effect of dispersion has also been used to explain features that we see in optical aurora. For example, *Samara and Michell* [2010] proposed that an alternative explanation to the common feature they observed in pulsating patches could be the result of higher-energy electrons arriving first followed by a greater flux of lower energy electrons, where the steep decrease in luminosity corresponds to the emptying of the loss cone. *Nishiyama et al.* [2014] suggested that the dispersion effect can explain why their ground-based data do not show any strong modulation with frequency higher than 3 Hz. The explanation being that the time difference between high-energy and low-energy electrons can easily fill the time gap of precipitating electron flux due to the quiet period between chorus elements. They further used the results of *Saito et al.* [2012] to explain that this effect seems to be larger on modulations generated by chorus elements with a repetition rate of 10 Hz compared to 3 Hz. *Saito et al.* [2012], however, explain their simulation results differently. Count rates of precipitated 1 MeV electrons detected at 100 km altitude show a pulse structure for 300 ms chorus element repetition, which is not seen for a 100 ms repetition. They explain that this is due to the precipitation of one chorus element overlapping the next for a time gap as short as 100 ms. Hence, it is not a time dispersion effect. In any case, it is clear that it is not straightforward to deduce the exact effect of dispersion in pulsating aurora. However, because the most probable time lag due to time dispersion is 0.6 s or less, and this is possibly further decreased by the effects of the pitch angle scattering mechanism, it is likely that the observed characteristics in the energy deposition are due to the generation mechanism and not a dispersion effect.

### 6.4. Observational Constraints on the Suggested Mechanisms

The auroral observational consequences of the most frequently mentioned theories of PA and how they fit within the observational constraints of PA patches are listed in Table 2. In the recent years, there has been an increase in the observations at the magnetic equator resulting in many studies relating different local plasma parameters (chorus waves, electron cyclotron harmonic waves, plasma densities, and ULF waves) and linking these to characteristics of PA. Below we will further discuss observational constraints set by the PA characteristics on the proposed theories and studies.

A significant result from our analysis is the highly variable on-time and the short off-times. This begs the fundamental question: What controls the variable on-time and off-time? In section 6.3 we found that the frequently cited theoretical candidates explaining the time-varying pitch angle scattering did not appear to provide likely explanations, although our findings fit better within the framework of the nonlinear relaxation oscillator, compared to the flow cyclotron maser. A similar conclusion can be drawn from a simple comparison to the expected on-time and off-time of the models. The nonlinear relaxation oscillator could give pulsation periods of order 3 to 30 s mainly determined by the strong diffusion time, the bounce period, or the loss cone filling time; most calculations yield values closest to the bounce period [*Davidson*, 1990]. If the latter is the case, it could possibly explain the preferred (typical) on-time of ∼3–5 s, which according to our event-specific calculations are close to 3.6 s for 10 keV electrons, 2.4 s for 20 keV electrons, and 1.2 s for 100 keV electrons. Interestingly, *Davidson and Chiu* [1991] later described a system that even for a purely sinusoidal driver could result in extremely complex electron precipitation pulsations. The measurable consequences could be trains of pulsations that begin between the pulses of a train that started several cycles earlier and a Fourier spectrum of mainly broad features rather than narrow peaks associated with periodic behavior. Both the periods and the variability of the pulsations that we observe could fit within this framework. The flow cyclotron maser explicitly includes the effect of the entry of new energetic electrons into a flux tube with enhanced cold plasma density which serves as a resonance cavity and deals with details of the wave-particle interactions which makes the theory easier to compare with observations [*Nemzek et al.*, 1995]. The on-time is determined by the nonlinear dynamics and is on the order of wave travel time between the ionospheres ∼1–5 s, whereas the off-time is nearly equal to the time of energetic electron accumulation in the duct, suggesting a repetition period of ∼5–30 s [*Demekhov and Trakhtengerts*, 1994]. These relatively short on-times do not fit our observations. In a later study *Tagirov et al.* [1999] explain observations of pulsating patches based on the flow cyclotron maser theory and an ionospheric feedback. In this process the relaxation time is about twice the buildup time and the whole cycle of the order 5–15 s. Our observations show weak evidence of a quicker buildup than decay but have a wider temporal range. The observations described in section 6.3 show an agreement with the nonlinear relaxation oscillator, while here the observation could be found in favor of a mechanism based on the flow cyclotron maser theory. Based on the expected temporal characteristics, our observations fit better within the framework of the nonlinear relaxation oscillator. However, this is mainly because the model predicts a large temporal variability, which also could make it fit to almost any observation.

The temporal characteristics of our observations can be compared to the chorus bursts which are found to correlate with plasma density depletions, but the agreement is not convincing. An additional factor suggested to play a dominant role in determining the on-off duration is temporal variations in the plasma density at the magnetic equator. Observations at geosynchronous orbit have revealed that quasiperiodic chorus waves can arise from changes in the plasma density probably attributed to changes in cold electron (less than a few eV) fluxes [*Li et al.*, 2011b, 2012]. A visual comparison of the temporal characteristics of the chorus magnetic wave amplitude is in good correlation with the density depletions by *Li et al.* [2011b] which show ∼10 s on-times and short off-times in one case but ∼15–30 s on-times and even longer off-times in the other case. Hence, the temporal characteristics of the wave bursts correlating with plasma density depletions can compare to our observations, but the agreement is not convincing.

Further, we find that the observation of ULF waves driving the pulsating aurora does not fit well within the observational constraints of pulsating auroral patches. Observations of low-energy ion precipitation by low-Earth orbit satellites have further suggested that the changes in the cold plasma in relation to pulsating auroral patches might owe its origin to the ion outflows from the ionosphere [*Liang et al.*, 2015], while other studies have related the growth of quasiperiodic chorus waves directly to the dynamics of ultralow-frequency (ULF) waves [*Li et al.*, 2011a; *Jaynes et al.*, 2015]. *Jaynes et al.* [2015] suggest that substorm-driven Pc4–Pc5 magnetospheric ULF pulsations (field line resonance as a result of a substorm injection) modulate chorus waves and thus are driver of pulsating particle precipitation. They link chorus modulations with ∼45 s to 1 min periods to ULF waves having periods closer to 2 min, thus occurring with twice the periodicity as the chorus. On the other hand, *Li et al.* [2011a] found a one-to-one variation in the ULF and chorus modulations, but also, they focused on the modulation of whistler mode waves by long-period compressional pulsations in the Pc4–Pc5 range and did not investigate individual chorus elements, but rather a group of chorus elements showing intensification over a time scale of tens of seconds to a few minutes. Thus, these time scales are too long to explain the on-times and off-times of pulsating auroral patches.

Our findings provide some strict observational constraints that any of the published theories must be able to explain. As such the wave-particle scattering mechanism where the off-time is the time it takes to reach the resonance threshold appears to be the leading candidate of PA patches. In Figure 13 we found that the off-time does not have a relation to the amount of energy that is deposited over the next on-time. This supports the wave-particle scattering mechanisms for which the scattered energy is not controlled by a preceding period of loading and where the off-time rather is the time it takes to reach the resonance threshold. For the flow cyclotron maser this is nearly equal to the time it takes for energetic electrons to accumulate in the flux tube and result in anisotropic velocity distribution. However, satellite observations of energetic electrons with large pitch angles are found to not change drastically between on and off stages [*Jaynes et al.*, 2013]. Further, *Nishiyama et al.* [2014] suggested that the on-off period of the pulsating aurora having ∼3 Hz modulations superimposed was likely due to ULF compressional waves creating temporal variations in the cold plasma density affecting the conditions for nonlinear wave growth and consequently switch strong pitch angle scattering on and off repeatedly. The off-time can then be the time it takes for the cold plasma to lower the resonance threshold. If the suggested ULF waves act as a quasiperiodic driver, we would expect the temporal range of the off-time to be relatively narrow, which is exactly what we observe. However, the ULF waves described in the above paragraph of course have too long periods to explain the 0.6 s median off-time of PA patches. In summary, the characteristics of the energy deposition indicates that the off-time is the time it takes to reach the resonance threshold, and the relative narrow temporal range of the off-time can fit with the idea that the latter is controlled by a driver which is relatively periodic. We realize that any of the suggested drivers can account for a lack of agreement with our results by claiming changes in the plasma properties (e.g., density) in the generation region. We, however, argue that such support is a patchy unsatisfactory argument.

Alternative drivers to the traditional idea of a pitch angle scattering by chorus or electron cyclotron harmonic waves likely do not fit within the observational constraint of pulsating auroral patches. The most recent suggestions are (1) Fermi-type accelerations accompanying earthward plasma flow [*Nakajima et al.*, 2012] and (2) near-Earth time-varying field-aligned potential [*Sato et al.*, 2002, 2004]. *Nakajima et al.* [2012] observed pulsating aurora which traces to the midtail equatorial plane region at a much further distance than most other studies on pulsating aurora. They find that the local plasma conditions are not favorable to the generation of whistler and ECH waves through the electron temperature anisotropy and suggest that field-aligned electrons are already generated by Fermi-type accelerations associated with earthward plasma flow at substorm onset. However, the pulsations of the field-aligned electrons are further suggested to be created by small pitch angle modulations by weak whistler mode waves, ECH waves, and/or magnetosonic waves, or alternatively Alfvén waves. *Sato et al.* [2002, 2004] observed nonconjugate east-west aligned arc type of pulsating aurora and suggested that they are driven by near-Earth time-varying field-aligned electric fields. These studies do not go into detail on the proposed driver; however, there are other studies that go into detail on how quasi-oscillatory variation of parallel electric fields can arise. *Pilipenko et al.* [1999] discuss how ion-acoustic instabilities can arise when densities of local field-aligned currents in the topside ionosphere reach the threshold for excitation of high-frequency turbulence. These are related to nightside substorm onset. Further, the ion-acoustic instability gives rise to quasi-oscillatory variations of parallel electric fields and consequently particle acceleration and precipitation. Alternatively, *Fedorov et al.* [2004] discuss how magnetospheric Alfvén waves penetrating into the auroral acceleration region can produce oscillatory variations of the field-aligned potential drop in the auroral topside ionosphere that may cause oscillatory frequency dependence of electron acceleration modulations in the range around fractions of a hertz. In conclusion, the recently suggested alternative drivers are not sufficiently mature to provide any observational predictions and thus we cannot discuss them in a quantitative manner. Further, they are likely to result in a quasi-oscillatory particle precipitation which do not fit well within the observational constraint of pulsating auroral patches.

## 7. Summary and Conclusions

We presented a careful study of PA patches using green line all-sky images obtained at 3.3 Hz. We identified six individual pulsating patches and extracted them using a contouring technique. This allowed us to derive objective quantitative parameters for each of the patches. These included patch duration (on-time and off-time), peak intensity, and integrated intensity. Further, we found how the temporal characteristics related to the characteristics of the energy deposition. Altogether, we provide a series of observational constraints on the suggested mechanisms.

The PA patches display a striking temporal variability. The distribution of on-times shows a large spread (2–21 s) but is clearly not scaleless indicating that there is a preferred (typical) on-time of ∼3–5 s. The distribution of the off-time is more confined (0–7 s) having a median of 0.6 s. This clearly states that the frequently used period or quasiperiod cannot sufficiently describe the temporal characteristics. On-time and off-time serve as a more accurate description of the observational constraint. We argue that the naming of pulsating aurora should be changed to fluctuating aurora since it is not periodic but rather erratic.

The temporal constraints do not appear to support the variations of the flow cyclotron maser theory but fit better within the framework of the nonlinear relaxation oscillator, largely because it predicts a temporal variability.

The energy deposition is found to be highly variable from pulse to pulse. The constraints set by the temporal characteristics in relation to the energy deposition seem to indicate the following: (1) The pulse on-time is a composition of higher-frequency pulsations, where the composition for some reason varies from pulse to pulse. (2) The off-time is not related to a loading but is related to the time it takes to lower the resonance threshold to start a new pulsation. (3) The slightly quicker buildup than decay of the energy deposition is opposite to the observational consequences that would be expected from the flow cyclotron maser (and backward wave oscillator), which is suggested to create the PA patches in particular.

It is clear that the suggested mechanisms and drivers of PA do not explain the observational constraints set by the PA patches in a satisfactory manner. Further, observations of PA patches and other types of PA with ground-based instruments are required to obtain a large set of observational constraints that any suggested driver or mechanism of PA must explain.

**Acknowledgments**

This study was supported by the Research Council of Norway under contract 223252/F50. The authors acknowledge the use of SuperMAG indices and all-sky imager data from the Multi-spectral Observatory of Sensitive EMCCDs (MOOSE). The SuperMAG indices were obtained freely from supermag.uib.no. MOOSE all-sky imager data were obtained from R. Michell and M. Samara. The data analyzed in this study are available upon request from the authors.